\documentclass[conference]{IEEEtran}
\makeatletter
\def\ps@headings{%
\def\@oddhead{\mbox{}\scriptsize\rightmark \hfil \thepage}%
\def\@evenhead{\scriptsize\thepage \hfil \leftmark\mbox{}}%
\def\@oddfoot{}%
\def\@evenfoot{}}
\makeatother
\IEEEoverridecommandlockouts

\usepackage{multirow}
\usepackage{array}
\usepackage{cite}
\usepackage{amsmath,amssymb,amsfonts}
\usepackage{algorithmic}
\usepackage{graphicx}
\usepackage{textcomp}
\def\BibTeX{{\rm B\kern-.05em{\sc i\kern-.025em b}\kern-.08em
    T\kern-.1667em\lower.7ex\hbox{E}\kern-.125emX}}

\usepackage{soul}

\ifCLASSOPTIONcompsoc
    \usepackage[caption=false, font=normalsize, labelfont=sf, textfont=sf]{subfig}
\else
\usepackage[caption=false, font=footnotesize]{subfig}
\fi

\usepackage{balance}

\usepackage[hyphens,spaces,obeyspaces]{url}
\ifodd 1
\usepackage{soul, xcolor}

\newcommand{\rev}[1]{{{\color{black} #1}}}

\newcommand{\shaun}[1]{{{\color{black} #1}}}

\fi

\title{Detecting ADS-B Spoofing Attacks using \\ Deep Neural Networks}

\author{Xuhang Ying$^{\ast}$, Joanna Mazer$^{\ast}$, Giuseppe Bernieri$^{\dagger}$, Mauro Conti$^{\dagger}$, Linda Bushnell$^{\ast}$, and Radha Poovendran$^{\ast}$\\
$^{\ast}$ Department of Electrical and Computer Engineering, University of Washington, Seattle, WA 98195. \\
$^{\dagger}$ Department of Mathematics, University of Padua, Padua, Italy. \\
Email: \{xhying, jmazer, lb2, rp3\}@uw.edu, \{bernieri, conti\}@math.unipd.it
}

\begin{document}

\maketitle

\begin{abstract}
The Automatic Dependent Surveillance-Broadcast (ADS-B) system is a key component of the Next Generation Air Transportation System (NextGen) that manages the increasingly congested airspace.
It provides accurate aircraft localization and efficient air traffic management and also improves the safety of billions of current and future passengers.
While the benefits of ADS-B are well known, the lack of basic security measures like encryption and authentication introduces various exploitable security vulnerabilities.
One practical threat is the ADS-B spoofing attack that targets the ADS-B ground station, in which the ground-based or aircraft-based attacker manipulates the International Civil Aviation Organization (ICAO) address (a unique identifier for each aircraft) in the ADS-B messages to fake the appearance of non-existent aircraft or masquerade as a trusted aircraft.
As a result, this attack can confuse the pilots or the air traffic control personnel and cause dangerous maneuvers.

In this paper, we introduce SODA -- a \shaun{two-stage} Deep Neural Network (DNN)-based \underline{s}p\underline{o}ofing \underline{d}etector for \underline{A}DS-B that consists of a message classifier and an aircraft classifier.
It allows a ground station to examine each incoming message based on the PHY-layer features (e.g., IQ samples and phases) and flag suspicious messages.
Our experimental results show that SODA detects ground-based spoofing attacks with a probability of 99.34\%, while having a very small false alarm rate (i.e., 0.43\%).
It outperforms other machine learning techniques such as XGBoost, Logistic Regression, and Support Vector Machine. 
It further identifies individual aircraft with an average F-score of 96.68\% and an accuracy of 96.66\%, with a significant improvement over the state-of-the-art detector. 




\end{abstract}

\section{Introduction}
Rapid growth in the air traffic expected for the coming years requires innovative applications and methodologies to guarantee efficiency of the transportation infrastructure and the safety for passengers and crew. 
In order to monitor and manage the increasingly congested airspace, the Automatic Dependent Surveillance-Broadcast (ADS-B) system is deployed as a key component of the Next Generation Air Transportation System \cite{nextgen}.
The airborne ADS-B OUT devices (transmitters) allow aircraft to periodically broadcast their identifications and current positions, determined from global navigation satellite systems, to ground stations or other aircraft over the 1090 MHz band.
\shaun{All aircraft will need to be equipped with ADS-B capabilities to be operated in the European and U.S. airspace by 2020, as per the EASA and FAA\footnote{EASA and FAA stand for the European Aviation Safety Agency and the Federal Aviation Administration, respectively.} regulations.}



Despite of the advantages of ADS-B, its security vulnerabilities have been widely acknowledged  \cite{sampigethaya2011security,mccallie2011security,strohmeier2015security,costin2012ghost,schafer2013experimental}.
\shaun{In particular, the source of the problem is the lack of basic security mechanisms,} such as encryption and authentication. 
As a result, a variety of wireless attacks can be launched against the ADS-B 
system, such as eavesdropping, jamming, message injection (spoofing), message deletion, and message modification, as summarized in  \cite{mccallie2011security}.
The above security vulnerabilities of ADS-B are further exacerbated by the prevalence of low-cost Software Defined Radios (SDRs) (e.g., USRP \cite{usrp}).
It has been demonstrated in \cite{schafer2013experimental,costin2012ghost} that the above attacks are easy and practically feasible to launch for a moderately sophisticated attacker with a SDR.

\begin{figure*}[t]
\centering
\includegraphics[width=0.9\textwidth]{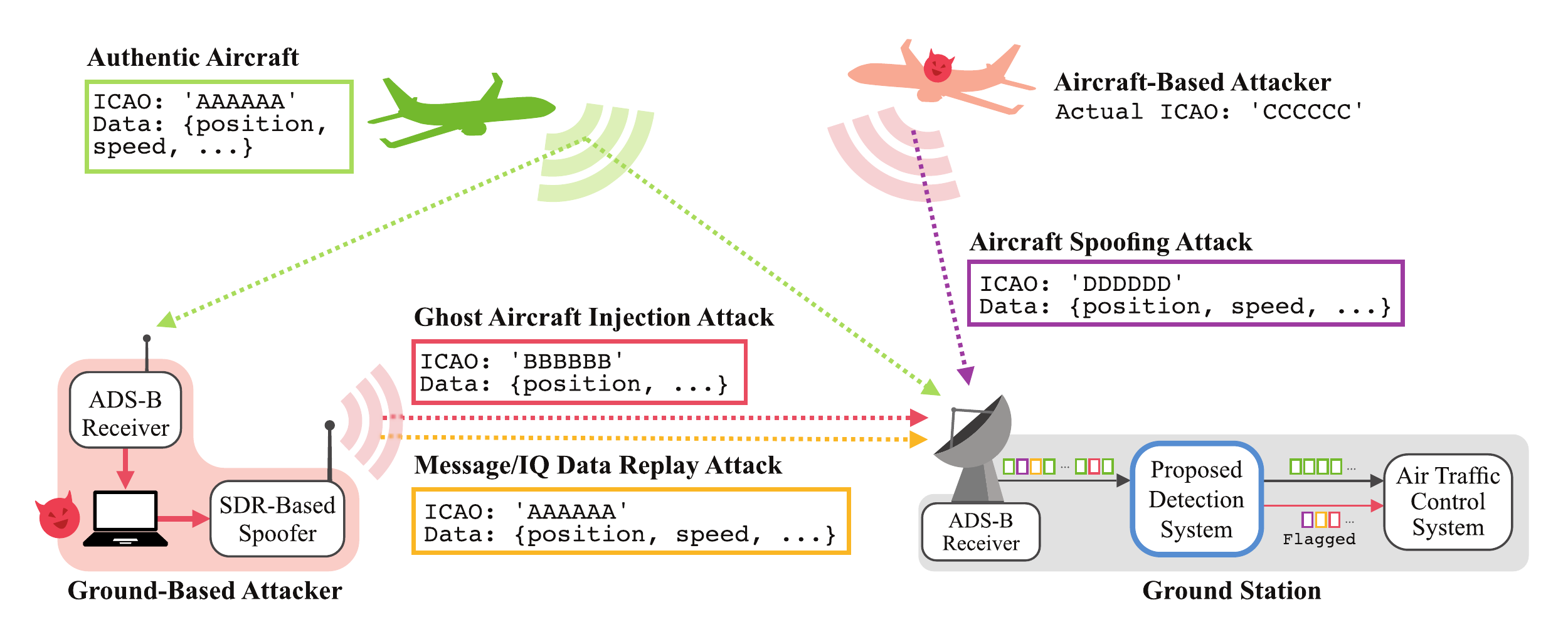}
\caption{Illustration of ADS-B message injection or spoofing attacks that target the ground station.
The ground-based attacker uses a SDR device to replay previously recorded ADS-B messages or transmit \rev{newly generated and correctly modulated fake} 
ADS-B messages, whereas an aircraft-based attacker modifies the ICAO address of transmitted messages and masquerades as a known or trusted aircraft.
SODA is deployed between the ADS-B receiver and the ATC system and flags suspicious messages based on the PHY-layer features using DNN. }
\label{fig:main_figure}
\end{figure*}

In this work, we focus on the ADS-B message injection or spoofing attacks \shaun{that target} a ground station.
As illustrated in Fig.~\ref{fig:main_figure}, the ADS-B receiver receives broadcast messages \shaun{broadcast by aircraft within signal range} and feeds them to the Air Traffic Control (ATC) system.
We classify the ADS-B spoofing attacks into two categories: ground-based and aircraft-based.
In the first attack, the ground-based attacker uses a SDR to retransmit previously recorded messages or transmit newly generated and correctly modulated fake messages, which are referred to as the replay attack and the ghost aircraft injection attack, respectively. 
The main objective is to fake the presence of  non-existent (ghost) aircraft and confuse the ATC system. 
In the second attack, an aircraft-based attacker modifies the 
\rev{ICAO} address 
in the ADS-B messages transmitted by the airborne ADS-B transponder, masquerading itself as a known or trusted aircraft to bypass the surveillance. 

Use of cryptographic primitives (e.g., Message Authentication Code) is one way to defend against ADS-B spoofing \cite{wesson2014can,finke2013enhancing,viggiano2010secure,schuchman2011automatic,feng2010data}, but it requires protocol modifications or overheads and \rev{shifts} 
the problems to the design of secure and scalable key management and distribution schemes.
Alternative non-cryptographic approaches aim to identify and exploit unique characteristics of ADS-B transponders for fingerprinting. 
In \cite{strohmeier2015passive}, Strohmeier \textit{et al.} analyzed message transmission patterns (inter-arrival time distribution) and identified distinct \rev{patterns} 
between commercial ADS-B transponder types and their implementations.
Nevertheless, an attacker with a SDR-based spoofer has the full control of inter-transmission times and can mimic the transmission pattern to evade detection.
A malicious aircraft can also masquerade as another aircraft equipped with an ADS-B transponder of a similar implementation (e.g., from the same manufacturer) to avoid detection. 
In \cite{leonardi2017air}, Leonardi {\em et al.} proposed to fingerprint aircraft through phase patterns and applied neural networks for multi-class classification. 
Due to the limited number of predefined classes for phase patterns (seven in total), there will be a significant number of aircraft in each class (e.g., 100+ aircraft in the class of non-coherent phase patterns), leaving a lot of freedom to the attacker.

In this work, we present SODA -- a Deep Neural Network (DNN)-based \underline{s}p\underline{o}ofing \underline{d}etector for \underline{A}DS-B. 
SODA takes PHY-layer features (e.g., IQ samples and phases) as input and performs two-stage detection to detect malicious messages and identify malicious aircraft.

\vspace{0.1cm}
\noindent \textbf{Contributions.} 
Throughout this paper, we make the following contributions:
\begin{itemize}
    \item We introduce SODA, a DNN-based detector that consists of a message classifier and an aircraft classifier to detect the ground-based and aircraft-based spoofing attacks.

    \item We build a SDR-based spoofer and conduct Over-The-Air (OTA) experiments in an anechoic chamber to emulate realistic ground-based spoofing attacks.
    Our experimental results show that the message classifier detects malicious messages from a ground-based attacker with a detection probability of $99.34\%$ and a negligible false alarm rate (i.e., $0.43\%$).
    It outperforms baseline machine learning techniques including XGBoost, Logistic Regression, and Support Vector Machine. 
    
    \item \shaun{We build an aircraft classifier that performs multi-class classification with each aircraft being its own class.
    Our experimental results show that a fine-tuned aircraft classifier classifies a total of 238 aircraft with an average F-score of $96.68\%$ and an accuracy of \shaun{$96.66\%$}, with a significant performance improvement over the state-of-the-art detector in \cite{leonardi2017air}.}

\end{itemize}


\noindent \textbf{Organization.} The remainder of this paper is organized as follows. 
Section~\ref{sec:related_work} reviews related work and Section~\ref{sec:background} provides relevant background on ADS-B and DNN. 
In Section~\ref{sec:system_adversary_model}, we describe our system and adversary models.
Section~\ref{proposed_scheme} presents the proposed detection system, and Section~\ref{sec:evaluation} presents the experimental evaluation. 
This study is concluded in Section~\ref{sec:conclusion}. 


    
\section{Related Work}\label{sec:related_work}
The security vulnerabilities of ADS-B have been analyzed in many studies \cite{sampigethaya2011security,mccallie2011security,strohmeier2015security,costin2012ghost,schafer2013experimental}.
One way to secure wireless ADS-B communication is through cryptographic measures \cite{wesson2014can,finke2013enhancing,viggiano2010secure,schuchman2011automatic,feng2010data}.
Finke \textit{et al.} compared various encryption schemes and supported a symmetric cipher using the FFX algorithm \cite{finke2013enhancing}. 
Nevertheless, it is acknowledged as a challenging task to perform secure key management and distribution.
More recently, Wesson \textit{et al.} proposed the use Public Key Infrastructure (PKI) and ECDSA (Elliptic Curve Digital Signature Algorithm) signatures as the cryptographic solution. 
In \cite{feng2010data}, Feng \textit{et al.} presented a PKI solution for ADS-B message authentication based on Elliptic Curve Cipher and X.509 certificates. 
But the proposed scheme would require additional messages to carry the signature and timestamps, which is hardly scalable in practice, and efficient certification distribution remains an open question.

Alternative non-cryptographic approaches can be deployed to mitigate the attacks on the ADS-B system.
In \cite{leonardi2017ads}, the authors evaluated ADS-B jamming attacks and proposed a solution based on PHY-layer signal separation. 
In \cite{schafer2015secure}, Sch{\"a}fer \textit{et al.} proposed secure track verification based on timing. 
In \cite{ghose2015verifying}, Ghose and Lazos exploited the Doppler spread phenomenon  to estimate actual velocities and crossed them with the claimed velocities.  
In \cite{schafer2016secure}, Sch{\"a}fer \textit{et al.} exploited the Doppler shift measurements from multiple ground stations (verifiers) to verify the motion of the aircraft (prover).
\shaun{While it was implicitly assumed that an attacker could modify the transmission frequency to mimic the Doppler shift and deceive a single ground station, we show that it could be a challenging task for the attacker, if the detector leverages the PHY-layer features for detection as SODA does.}
In \cite{kim2017ads}, the authors proposed an enhanced ADS-B protocol called ADS-BT, which embeds transmit timestamps in the ADS-B messages, but such approach would require protocol modifications.

In the more recent study in \cite{leonardi2017air}, the authors identified 7 classes of phase patterns and exploited them for aircraft classification. 
They applied a neural network (one hidden layer with $10$ nodes) and reported an accuracy of $91.4\%$ for the 7-class classification task.
While the results are promising, the effectiveness of the proposed detector is limited by the number of classes.
In other words, more aircraft in one class imply the greater freedom for an attacker to launch the aircraft spoofing attack.
In the ideal case, there should be only a single aircraft in each class. 
Moreover, a \shaun{ground-based} attacker in possession of a SDR-based spoofer can also mimic the phase pattern to evade the detection. 

In this work, we develop SODA -- a DNN-based two-stage spoofing detector. 
In the first stage, the message classifier examines each incoming message and labels it as malicious or non-malicious. 
For those that are considered non-malicious in the first stage, they are further processed by the aircraft classifier, which predicts the ICAO address based on the PHY-layer features and compares the prediction against the claimed address to detect aircraft spoofing attacks.
\shaun{Our experimental results demonstrate the effectiveness of the proposed SODA with a detection probability of $99.34\%$ in the first stage and an accuracy of $96.66\%$ for a total of 238 aircraft (with each aircraft being its own class) in the second stage. }
\vspace{0.5cm}
\section{Background}\label{sec:background}
In this section, we provide relevant background on the ADS-B protocol (Section~\ref{sec:adsb_background}) and a brief introduction to DNN (Section~\ref{sec:ddn_background}).


\subsection{ADS-B Background}
\label{sec:adsb_background}

Aircraft use the ADS-B technology to periodically transmit their identifications (ICAO addresses) and status data (e.g. position, speed, heading, etc.) to ground stations and nearby aircraft.
There are two types of ADS-B devices: 1) ADS-B OUT devices for transmitting broadcasts to ADS-B receivers, and 2) ADS-B IN devices for receiving broadcasts, weather data, and ATC reports.
The data link standards for ADS-B are the 987 MHz Universal Access Transceiver (UAT) and the 1090 MHz Extended Squitter (1090ES).
The former was created specifically for aviation services such as ADS-B and requires new hardware, whereas the later integrates the ADS-B function to traditional Mode S transponders. 
In this work, we focus on the commercially used 1090ES data link.

\begin{figure}[t]
\centering
\includegraphics[trim={0 0cm 0 -.3cm},clip,width=1\columnwidth]{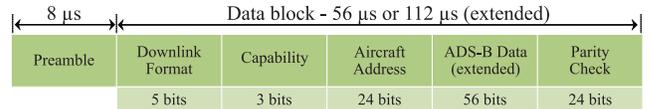}
\caption{Illustration of ADS-B message structure. Each message consists of a 8 $\mu$s preamble and a $56~\mu$s or $112~\mu$s data block. }
\label{fig:ads_b_message_structure}
\end{figure}

Fig.~\ref{fig:ads_b_message_structure} illustrates the ADS-B message structure.
Each message contains a 8 $\mu$s preamble for synchronization and a 56-bit (short) or 112-bit (extended) data block.
The first 5 bits of the data block contain the downlink format (the message type).
The subsequent 3-bit capability field serves as an additional identifier. 
The 24-bit ICAO address is a unique identifier issued to each 
aircraft by the ICAO. 
The 56-bit extended ADS-B data field contains surveillance information such as identification, position, velocity, and emergency codes. 
The last field is a 24-bit parity check for receivers to validate the correctness of the preceding message.
Within the 56-bit data field, the first 5 bits represent the Type Code (TC), indicating the type of information contained in the subsequent bits. 
While there are a total of 31 TCs, we focus on messages that include airborne position data with barometric altitude (TCs 9-18), and airborne velocities (TC 19). 

ADS-B messages are transmitted every $0.5$ s on average using the Pulse Position Modulation (PPM) scheme. 
Since the data rate is 1 Mbps, the total duration is 120 $\mu$s for an extended ADS-B message (including the preamble).

\subsection{Deep Neural Network}
\label{sec:ddn_background}
DNN is a widely used supervised learning technique, which requires a large amount of labeled training data. 
It consists of input, hidden, and output layers.
Each layer consists of a certain amount of nodes (neurons). 
The number of input nodes is equal to the number of features of a sample, and the number of output nodes usually correspond to the number of labels or classes. 
The number of hidden layers and nodes are considered as tuneable hyper-parameters.


In a feed-forward DNN with multiple hidden layers, with $j$ hidden nodes, each node maps its input, $x_j$, to the output state $y_j$, through an activation function. In Eq.~(\ref{eq1}), $x_j$ is equal to $b_j$, the bias of node \(j\), plus the sum of \(y_iw_{ij}\), over all \(i\) nodes in the preceding layer, where \(y_i\) is the output state of node \(i\) in the previous layer, and \(w_{ij}\) is the connecting weight from node \(i\) to node \(j\). 
The sigmoid function 
is often used in binary classification models, whereas the softmax function 
is used for multi-class classification problems, and returns the probability of each class, \(p_j\), over \(k\) total classes.
\begin{equation} \label{eq1}
x_j = b_j + \sum_{i}y_iw_{ij}.
\end{equation}



A cost function (e.g., cross entropy) is defined to measure the discrepancy between the target outputs and the actual outputs produced for each training sample.
The derivatives of the cost function are then back-propagated throughout the network to update the weights, thus training the DNN classifier.
In order to increase the efficiency of back-propagation on large training sets, mini-batches, i.e., small random set of training cases, are used in place of the entire training data. 
The term epoch refers to a single forward and back-propagation of the training data. 
More details on DNN are available in literature such as \cite{hinton2012deep}.


\section{System and Adversary Model}\label{sec:system_adversary_model}
In this section, we present our system model (Section~\ref{sec:system_model}) and our adversary model (Section~\ref{sec:adversary_model})  for ADS-B spoofing attacks against a ground station.  

\subsection{System Model}
\label{sec:system_model}
We consider a ground station equipped with an ADS-B receiver that has access to the raw IQ data.
We assume that the ground station has sufficient resources (e.g., processing power or storage) for various machine learning tasks. 
We also assume that the ground station has a database of previously received authentic ADS-B messages (including both message contents and IQ data) for each aircraft, which can be used to train a supervised learning-based detector.
In practice, such a database may be created by a ground station equipped with secondary surveillance systems (e.g., the radar surveillance system) such that the authenticity of each message can be cross-validated. 
This database can be later shared among other ground stations.
Attack messages can be collected by emulating the spoofing attacks using various SDR-based spoofers.
While a database created by a single ground station tends to be regional and does not cover all aircraft, multiple ground stations can collaborate and build a nation-wide dataset offline. 

\subsection{Adversary Model}
\label{sec:adversary_model}
As illustrated in Fig.~\ref{fig:main_figure}, we consider two types of attackers: 1) a \shaun{ground-based} attacker that uses a low-cost SDR-based spoofer, and 2) an \shaun{aircraft-based} attacker that uses its ADS-B transponder with modified ICAO address.
\shaun{While it is possible for an onboard attacker to transmit spoofed messages with the SDR-based spoofer from the aircraft, the signal may not reach the ground station due to the low transmit power of SDRs.
For example, the output power of USRP B210/B200 is +20 dBm (100 mW) max \cite{usrp}, whereas airborne ADS-B transponders are required to have a minimum transmission power of 75 W for smaller aircraft and 125 W or 200 W for larger aircraft \cite{rtca-260b}.
Spoofing attacks launched by an on-board attacker against the aircraft is beyond the scope of this paper.}

We assume that attackers have full knowledge of the ADS-B protocol and the location of the targeted ground station.
Since SDR devices are highly flexible and configurable through software, the attacker will \rev{have control over} 
signal characteristics (e.g., power level, phase, transmitting frequency). 
In particular, we assume that the attacker is aware of the Doppler effect \rev{(caused by aircraft movements)} and mimic the Doppler shift by shifting the transmitting frequency of the generated signal (or equivalently, by modifying the phases of the transmitted IQ samples). 
In contrast, commercial ADS-B OUT devices have custom circuits and do not provide the flexibility for users to access its physical layer or modify the signal characteristics. 

Depending on the way the spoofed messages are generated, we further divide the ADS-B spoofing attacks into three types: 1) message or IQ data replay attack, 2) ghost aircraft injection attack, and 3) aircraft spoofing attack.
The first two attacks can only be launched by a ground-based attacker, while the last one is launched by an aircraft-based attacker. 
Here we consider the replay attack as a special case of the spoofing attack, in which the attacker transmits previously recorded messages pretending to be the legitimate aircraft, without bothering to manipulate the message contents. 
We now describe each attack in detail.


\vspace{0.1cm}
\noindent \textbf{Message or IQ data replay attack.} 
In this attack, the ground-based attacker records the messages contents or the IQ data of the received authentic ADS-B messages using the SDR device, and then transmits the same messages at a later time without changing the message contents. 
Compared to the message replay attack, the IQ data replay attack is much stealthier, as the recorded IQ data incorporates a lot of information about the Doppler effect, the transmitter characteristics (e.g., carrier frequency offset), and the channel characteristics (e.g., multi-path and fading effects), which is difficult to mimic otherwise. 

\vspace{0.1cm}
\noindent \textbf{Ghost aircraft injection attack.} 
Unlike the replay attacks, the ground-based attacker transmits fake ADS-B messages with arbitrary contents of its choice using the SDR device. 
In particular, the attacker can simulate the trajectories of non-existent aircraft and generate corresponding ADS-B messages with carefully chosen Doppler shifts, causing ghost aircrafts to appear on the console of the ground station.

\vspace{0.1cm}
\noindent \textbf{Aircraft spoofing attack.} In this attack, an aircraft-based attacker (malicious aircraft) attempts to masquerade as a known or trusted aircraft by spoofing the ICAO address and hide its true identity. 
Since the aircraft is physically present, the masquerading attack will not be detected even if the secondary radar surveillance system is deployed.

\section{Proposed System -- SODA}
\label{proposed_scheme}
In this section, we first describe the architecture of SODA (Section~\ref{sec:system_architecture}).
We then present the DNN-based message classifier (Section~\ref{sec:message_classification}) and aircraft classifier (Section~\ref{sec:aircraft_classification}).

\subsection{System Architecture}
\label{sec:system_architecture}
\begin{figure}[t]
\centering
\includegraphics[trim={0 0cm 0 -0.3cm},clip,width=1\columnwidth]{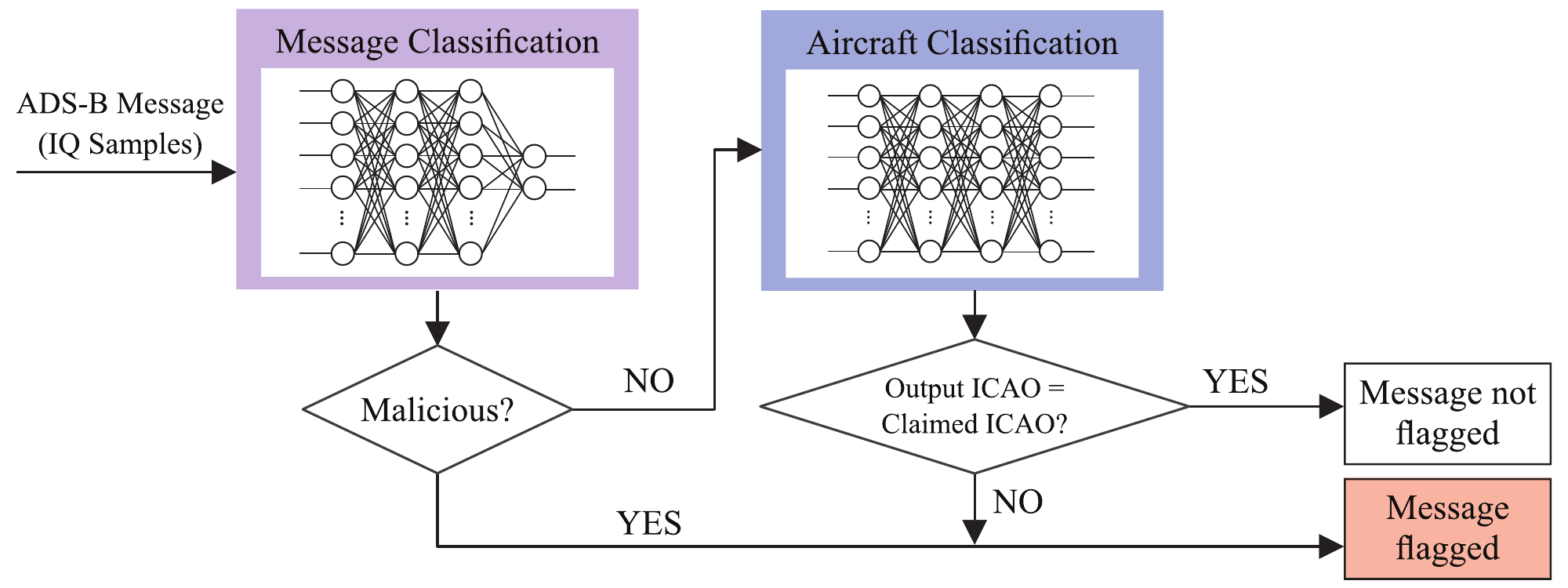}
\caption{Illustration of SODA Architecture.
The message classifier determines whether the ADS-B message is malicious or not so as to detect ground-based spoofing attacks. 
A non-malicious message will be further passed into the aircraft classifer. 
By comparing the output ICAO address with the claimed ICAO address, the detector can detect the aircraft-based spoofing attack.
}
\label{fig:system_architecture}
\end{figure}

Fig.~\ref{fig:system_architecture} illustrates the architecture of SODA, which consists of a message classifier and an aircraft classifier.
The message classifier decides whether the message is malicious or not.
If a message is considered non-malicious, it means that it is not transmitted by the SDR-based spoofer, but it may come from a malicious aircraft instead of the legitimate transmitter as indicated by the ICAO address in the message.
Hence, the aircraft classifier aims to further determine the \shaun{transmitting aircraft} of the message and compare the output ICAO address against the claimed ICAO address to detect the aircraft spoofing attack.

Since both message and aircraft classifiers are based on DNN, they need to be trained on a large labeled dataset. 
At runtime, the SODA takes each incoming ADS-B message (IQ data) as input and decides whether this message is suspicious or not. 
Since the SODA is a passive detection system that only flags possibly spoofed messages, it is up to the ATC system to take further actions (e.g., drop the flagged messages or display the ghost aircraft on the console). 
In the rest of this section, we describe the two stages in more detail. 

\subsection{Message Classification}
\label{sec:message_classification}
Message classification is \rev{considered as} 
a binary classification problem.
In SODA, the raw IQ samples are used as features.
Since each message lasts $120~\mu$s, a sampling rate of $R$ MHz will produce $240 R$ interleaving IQ samples. 

\begin{figure}[t]
\centering
\includegraphics[trim={0 0cm 0 0},clip,width=1\columnwidth]{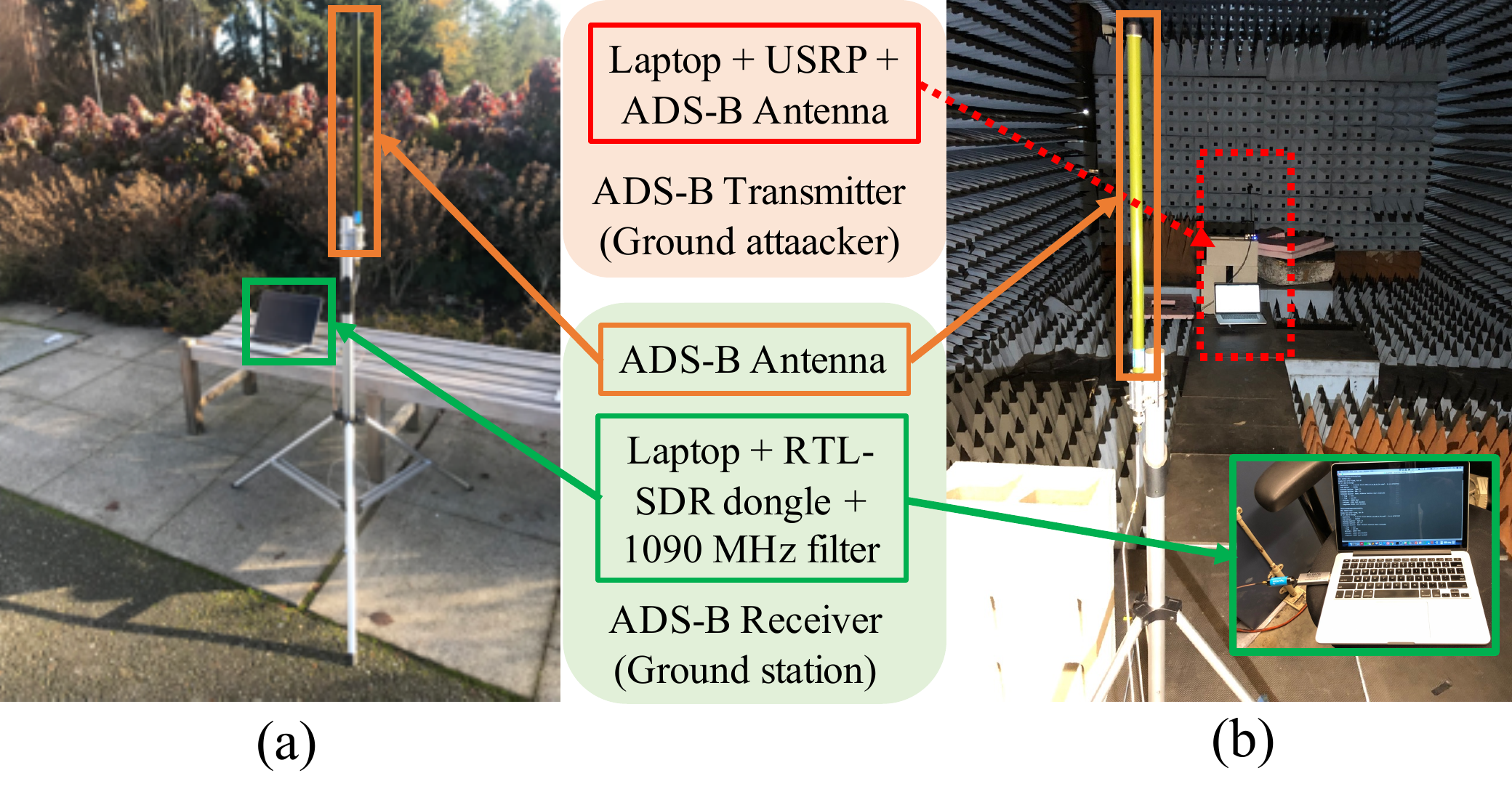}
\caption{Illustration of data collection using the constructed ADS-B receiver and SDR-based spoofer. (a) The ADS-B receiver collects authentic messages in an open area at the coordinates of $(47.6534, -122.3076)$. (b) The same ADS-B receiver receives malicious messages transmitted by the SDR-based spoofer in an anechoic chamber.}
\label{fig:hardware_setup}
\end{figure}

\vspace{0.1cm}
\noindent \textbf{Data collection.} 
To successfully train the message classifier, a large amount of realistic ground-based spoofing attack data is needed.
As illustrated in Fig.~\ref{fig:hardware_setup}(a), we constructed an ADS-B receiver (the ground station) that consists of a laptop, a RTL-SDR dongle \cite{rtlsdr}, a 1090 MHz bandpass filter, and an ADS-B antenna to collect real ADS-B messages in an open area. 
We then built a SDR-based ADS-B spoofer consisting of a USRP B210 \cite{usrp} and a 5-dBi 1090 MHz antenna (Fig.~\ref{fig:hardware_setup}(b)) to emulate ADS-B spoofing attacks. 

For the message replay attack, \rev{the spoofer transmits correctly modulated ADS-B signals using the message contents decoded from authentic ADS-B messages.}
For the IQ data replay attack, the spoofer simply replays the IQ data of the recorded ADS-B messages.
This represents the worst-case IQ data replay attack, because the replayed IQ data directly comes from the ground station, while in practice the attacker would record the IQ data using different hardware at a different location.
In the ghost aircraft injection attack, a number of aircraft are simulated with random ICAO addresses, velocities, altitudes, and headings.
For simplicity, we assume that each aircraft broadcasts an ADS-B message every second.

In order to simulate different received signal strengths of messages from the same aircraft when it is approaching or flying away from the receiver,  attack messages are transmitted multiple times with different USRP gains. 
Since the attacker may be aware of the Doppler effect and the existence of carrier frequency offsets, the calculated or random Doppler shifts and frequency offsets are incorporated into the transmitted attack messages.
More details are provided in Section~\ref{sec:evaluation}.

\begin{figure}[t]
\centering
\subfloat[\label{fig:example_authentic}]{
\includegraphics[trim={.4cm .3cm 0.5cm .5cm},clip,width=.48\columnwidth]{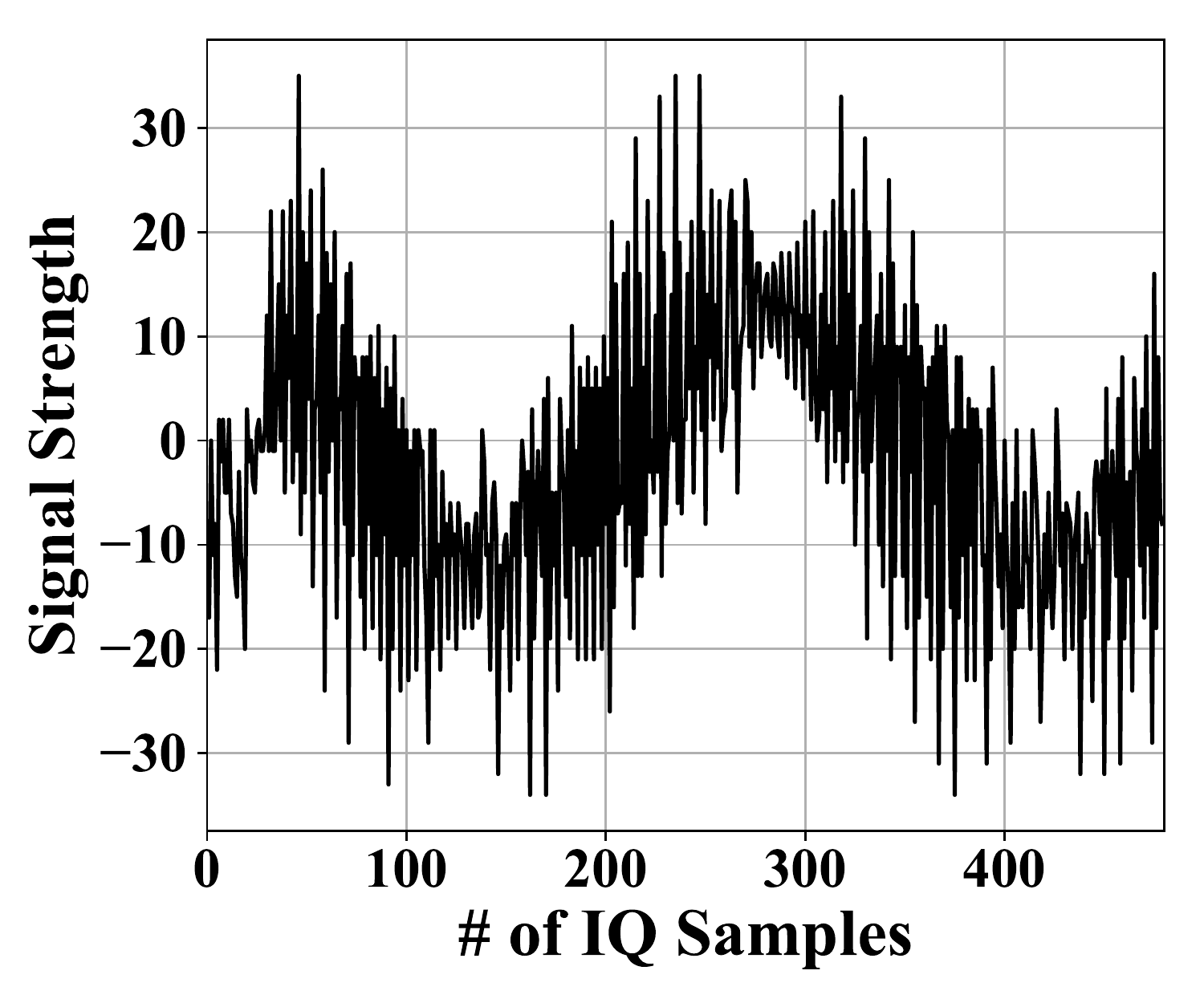}}
\hfill
\subfloat[\label{fig:example_message_replay}]{
\includegraphics[trim={.4cm .3cm 0.5cm .5cm},clip,width=.48\columnwidth]{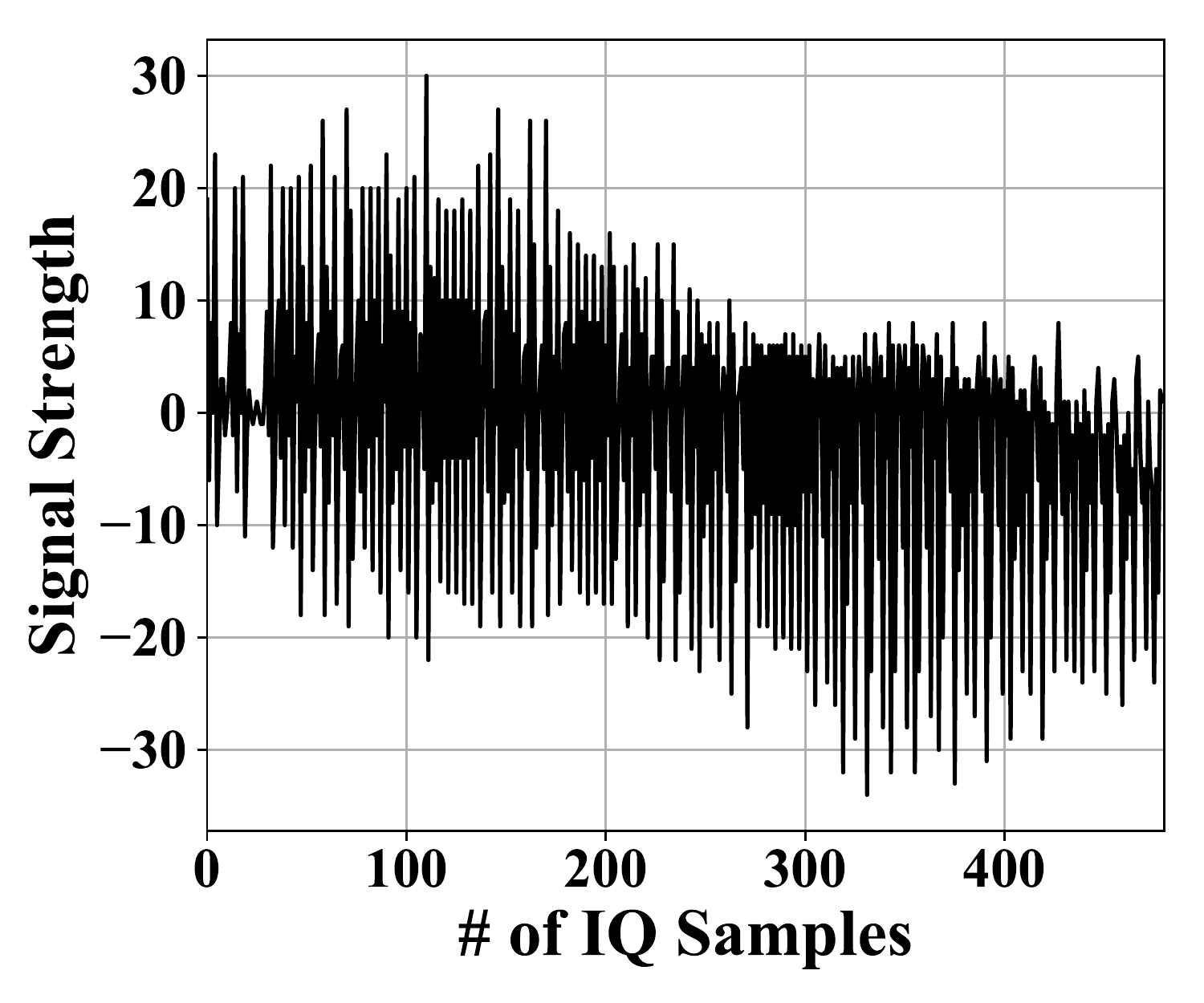}}
\hfill
\subfloat[\label{fig:example_iq_replay}]{
\includegraphics[trim={.4cm .3cm 0.5cm .5cm},clip,width=.48\columnwidth]{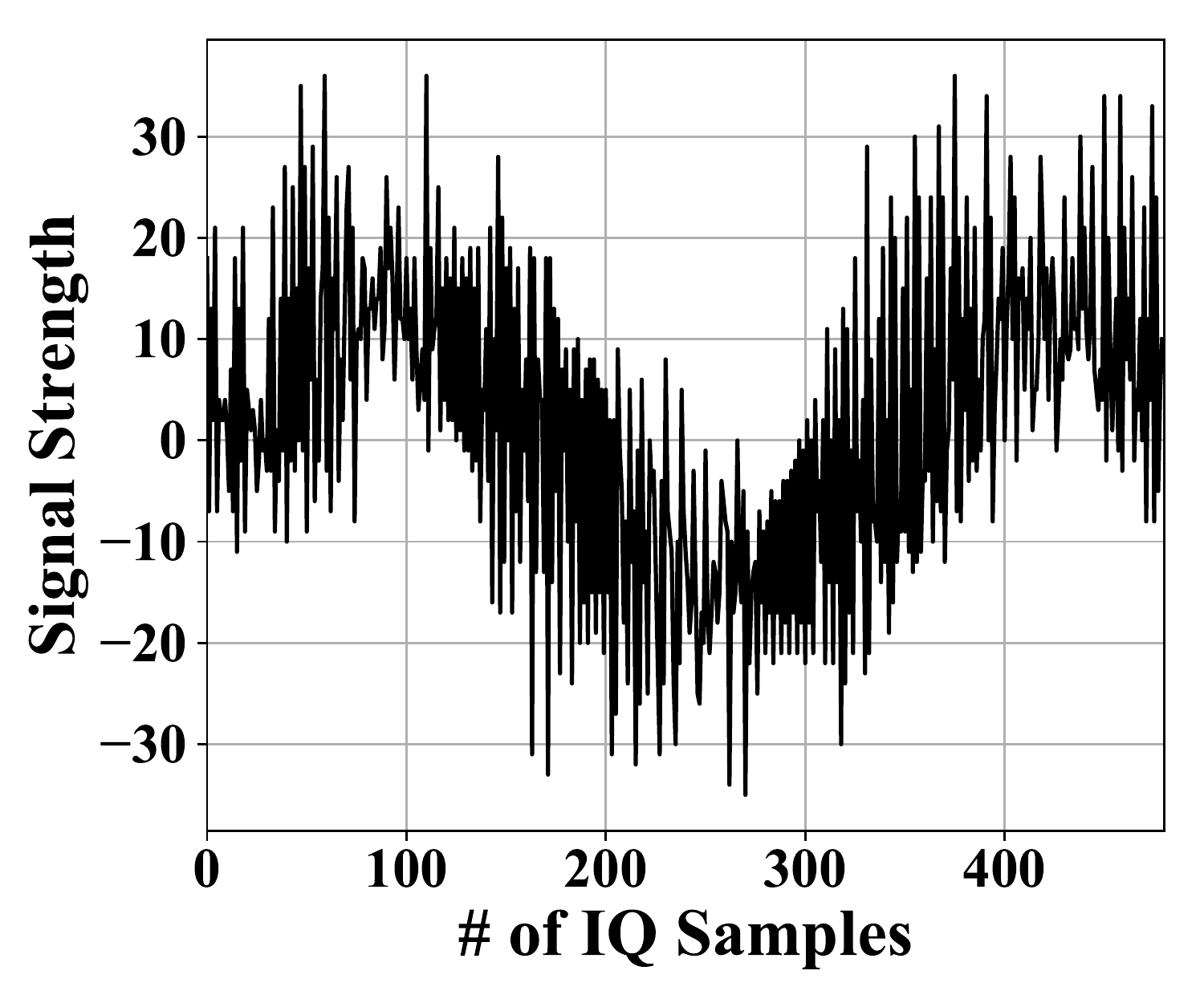}}
\hfill
\subfloat[\label{fig:example_ghost}]{
\includegraphics[trim={.4cm .3cm 0.5cm .5cm},clip,width=.48\columnwidth]{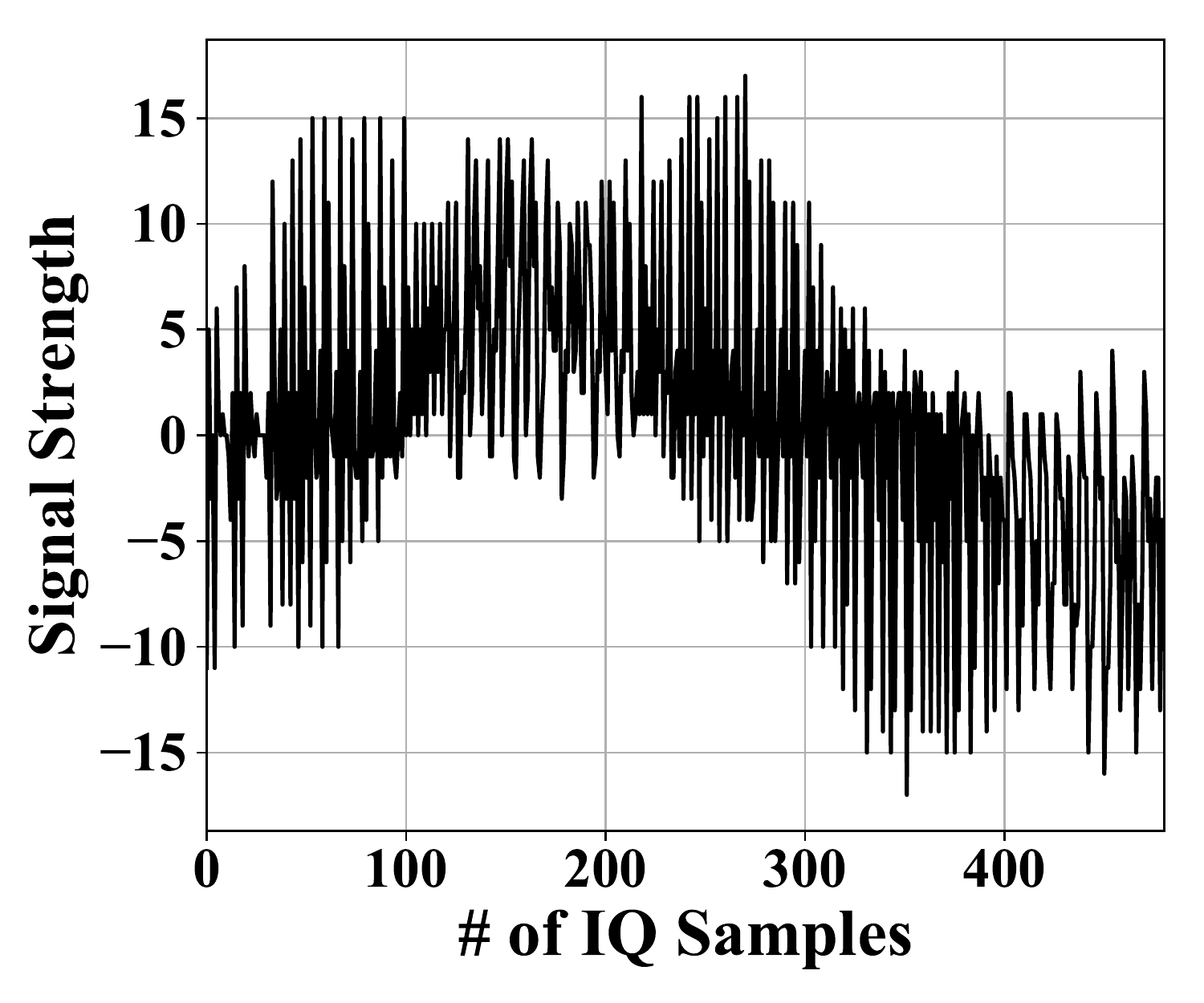}}
\caption{Examples of IQ samples of received ADS-B messages under different \rev{ground-based} ADS-B spoofing attacks. (a) No attack (authentic message). (b) Message replay attack. (c) IQ data replay attack. (d) Ghost aircraft injection attack. }
\label{fig:example_message_comparison}
\end{figure}

\vspace{0.1cm}
\noindent \textbf{Message classification example.} Fig.~\ref{fig:example_message_comparison} illustrates authentic and malicious messages under different attacks.
\rev{Recall that both message and IQ data replay attacks reuse the original message contents, whereas the ghost aircraft injection attack generates fake message contents.
Comparing Fig.~\ref{fig:example_message_comparison}(b) with Fig.~\ref{fig:example_message_comparison}(a), we notice that despite of the same message contents, the message replay attack leads to very different IQ samples. 
In contrast, the IQ data replay attack (Fig.~\ref{fig:example_message_comparison}(c)) can produce a very similar but shifted phase pattern, as compared to the original message (Fig.~\ref{fig:example_message_comparison}(a)).}



\subsection{Aircraft Classification} \label{sec:aircraft_classification}
The aircraft classifier predicts the source ICAO address of the received ADS-B message and compares it against the claimed ICAO address. 
Since there are many aircraft that transmit ADS-B messages, it then becomes a multi-class classification problem.
It is worth noting that through aircraft classification, the SODA can not only detect the aircraft spoofing attack, but also identify the masquerading aircraft.


\vspace{0.1cm}
\noindent \textbf{Features.}
Unlike message classification, the aircraft classifier does not use IQ samples or their magnitudes that encode the (possibly spoofed) ICAO address as features in order to avoid being deceived. 
Instead, it uses the phases that are independent of the claimed ICAO address as features. 
The phase of the $k$-th pair of IQ samples is computed as $$\phi[k] = \tan^{-1} \left(\frac{x_q[k]}{x_i[k]}\right).$$
From the communication theory \cite{madhow2008fundamentals}, we know that the phases encode the information about the TX and RX carrier frequency offsets and the Doppler shift.
To see why, denote the passband ADS-B signal as 
$$x_p(t)=\text{Re} \left\{ \sqrt{2}x(t)e^{j 2\pi f_c t} \right\},$$
where $x(t)=x_i(t) + jx_q (t)$ is the complex baseband signal and $f_c$ is the carrier frequency ($1090$ MHz for ADS-B signals).
With a carrier frequency offset of $\Delta f$ and a phase offset $\Delta \phi$, the resulting signal becomes
\begin{align*}
\tilde{x}_p(t) &= \text{Re} \left\{ \sqrt{2}x(t)e^{j 2\pi (f_c+\Delta f) t + \Delta \phi} \right\}\\
&= \text{Re} \left\{ \sqrt{2} \left(x(t) e^{j \phi(t)} \right) e^{j 2\pi f_c t} \right\},
\end{align*}
where $\phi(t) = 2\pi \Delta f t + \Delta \phi$. 
Hence, the rate of the change in the phase indicates the carrier frequency offset, which is a sum of TX/RX frequency offsets and the Doppler shift\footnote{Since the ADS-B signal is a narrowband signal, the impact of the Doppler effect is equivalent to shifting the carrier frequency by a certain amount.} and also affected by the propagation channel. 
\rev{As a result, the phases encode rich information and thus can be used as features for classification.}

\begin{figure}[t]
\centering
\includegraphics[width=1\columnwidth]{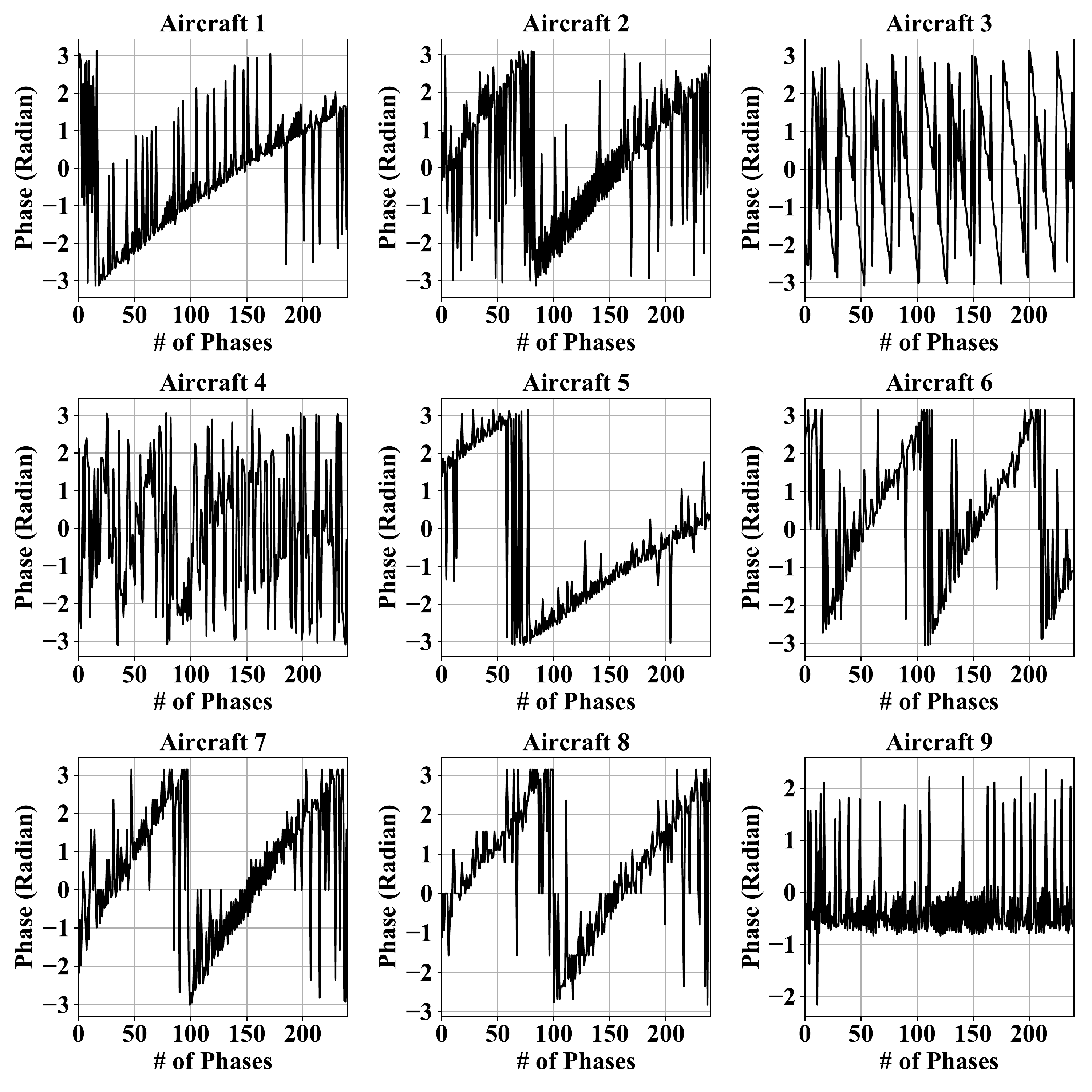}
\caption{Comparison of message phase patterns of different aircraft. }
\label{fig:example_aircraft_comparison_phase_over_time}
\end{figure}

\vspace{0.1cm}
\noindent \textbf{Aircraft classification example.}
A comparison of phase patterns of received messages from different aircraft is provided in Fig.~\ref{fig:example_aircraft_comparison_phase_over_time}. 
On one hand, we observe that aircraft \rev{could} 
have distinguishable phase patterns from \rev{each other}, which could be learned by the DNN-based aircraft classifier. 
On the other hand, it is \rev{possible} 
for two aircraft to have a similar phase pattern, such as Aircraft 7 and 8, which \rev{might} cause potential misclassifications. 

\section{Experimental Evaluation}
\label{sec:evaluation}
In this section, we describe the procedure for data collection and then evaluate the performance of both message and aircraft classifiers of SODA \rev{implemented using} Keras \cite{chollet2015keras} using real-world datasets. 

\subsection{Datasets}
The ADS-B receiver and SDR-based spoofer in Fig.~\ref{fig:hardware_setup} were used for data collection. 
The ADS-B antenna was mounted on a tripod and the overall height is about 1.5 meters.
A modified version of dump1090 \cite{dump1090} was used to collect IQ samples with a gain of 49.6 and a sampling rate of 2 MHz.
Both wired and OTA experiments were performed to collect realistic spoofing attack data. 
The former were conducted by directly wiring the transmitter and receiver. 
The latter were performed in an anechoic chamber located in the basement level of a building
to avoid possible signal emission.

\vspace{0.1cm}
\noindent \textbf{Datasets for message classification.}
Table~\ref{table:datasets} summarizes the collected datasets.
A total of 18675 authentic ADS-B messages (extended, downlink format of 17) were collected in an open area within half an hour. 
The first 10000 messages were used for message and IQ data replay attacks. 
A total of $20$ ghost aircraft were simulated with a velocity randomly drawn from a normal distribution $N(230 \text{ m/s}, 10 \text{ m/s})$.
The altitude follows $N(9000 \text{ m}, 500 \text{ m})$.
The headings (in degree) are drawn from a uniform distribution $U[0, 360)$. 

The Doppler shift $\Delta f_d$ and the carrier frequency offset $\Delta f_c$ are simulated in four different cases: i) $\Delta f_d = \Delta f_c = 0 $, ii) calculated $\Delta f_d$ and $\Delta f_c=0$, iii) calculated $\Delta f_d$ and random $\Delta f_c$, iv) random $\Delta f_d$ and $\Delta f_c=0$, and v) random $\Delta f_d$ and random $\Delta f_c$. 
The distributions of random $\Delta f_d$ and $\Delta f_d$ are $U[-1\text{ kHz}, + 1 \text{ kHz}]$ and $U[-10\text{ kHz}, + 10 \text{ kHz}]$, respectively.

\begin{table}[t]
\caption{Collected datasets for message classification.}\label{table:datasets}
\centering
\begin{tabular}{|c|c|c|c|c|} 
\hline
Dataset & Label & Wired & OTA & Total \\ \hline
Authentic messages & A0 & N/A & 18675 & 18675 \\ \hline
Message replay attack & A1 & 8101 & 8318 & 16419 \\ \hline
IQ data replay attack & A2 & 5961 & 5806 & 11767 \\ \hline
Ghost aircraft injection attack &  A3 & 8770 & 8832 & 17602 \\ \hline
\end{tabular}
\normalsize
\end{table}

\vspace{0.1cm}
\noindent \textbf{Datasets for aircraft classification.} 
In order to provide sufficient amount of data for aircraft classification, we collected additional data using the constructed ADS-B receiver on the rooftop of a building for two consecutive days. 
There are a total of 300,152 and 326,122 extended ADS-B messages from 157 and 210 aircraft in two days, respectively. 




\begin{table}[t!]
\centering
\caption{Basic DNN setup for SODA}
\label{table:DNN_parameters}
\begin{tabular}{|c|c|}
\hline
Description & Value \\ \hline
Input Layer & Number of features \\ \hline
Output Layer & Number of classes \\ \hline
Weight Initialization & Xavier normal \\ \hline
Weight Regularizer & L2 \\ \hline
Activation Function & ReLU \\ \hline
Output Activation Function & Softmax \\ \hline
Cost Function & Cross entropy \\ \hline
Optimizer & Adam \\ \hline
Epochs, Batch Size & 50, 32 \\ \hline
\end{tabular}
\end{table}

\subsection{Performance of Message Classification}
\label{sec:message_classification_performance}
In this experiment, we evaluate the performance of the DNN-based message classifier and compare it against three baseline classifiers: XGBoost \cite{chen2016xgboost}, Logistic Regression (LR) \cite{hosmer2013applied}, and Support Vector Machine (SVM) \cite{suykens1999least}.

\vspace{0.1cm}
\noindent \textbf{Setup.}
We first consider all attacks and split the dataset into 60\%, 20\%, and 20\% for training, validation, and testing, respectively.
The DNN parameters are summarized in Table~\ref{table:DNN_parameters}. 
We consider the following three DNN models:
\begin{enumerate}
    \item D1: one hidden layer with 128 nodes,
    \item D2: one hidden layer with 256 nodes, and
    \item D3: two hidden layers with 128 nodes per layer. 
\end{enumerate}
When training XGBoost/LR/SVM, the attack dataset is subsampled so that the attack and authentic datasets have the same number of messages, avoiding data imbalance.

\vspace{0.1cm}
\noindent \textbf{Metrics.}
Since each classifier acts as a detector, we adopt the following two metrics:
\begin{enumerate}
    \item \textit{Detection probability} (denoted as $P_d$): the percentage of malicious messages classified as malicious, and 
    
    \item \textit{False alarm probability} (denoted as $P_{fa}$): the percentage of authentic messages  classified as malicious.
\end{enumerate}
In practice, a classifier that achieves a higher $P_d$ and a lower $P_{fa}$ (that is within a given limit) is preferred.


\vspace{0.1cm}
\noindent \textbf{Results.}
The results are provided in Table~\ref{table:model_comparison}. 
We first observe that a simple DNN model such as D1 can already achieve very good performance with $P_d = 99.39\%$ and $P_{fa}=2.60\%$, as compared to other machine learning techniques.
In particular, LR and SVM do not perform well in this task, and XGBoost only achieves $P_d$ of $78.37\%$ with $P_{fa}=5.17\%$.
If we increase the complexity of the DNN model by doubling the number of nodes, the message classifier (D2) achieves $P_d = 99.50\%$ and $P_{fa}=2.09\%$.  
Interestingly, if the hidden nodes of D2 are split equally into two hidden layers, the resulting classifier (D3) can further reduce $P_{fa}$ to $0.43\%$, while keeping $P_d=99.34\%$.


\begin{table}[t]
\centering
\caption{Comparison of different models for message classification}
\label{table:model_comparison}
\begin{tabular}{|c|c|c|}
\hline
Model & $P_d$ & $P_{fa}$  \\ \hline
D1 & 99.39\% & 2.60\%  \\ \hline
D2 & 99.50\% & 2.09\%  \\ \hline
D3 & 99.34\% & 0.43\%  \\ \hline
XGBoost & 78.37\% & 5.17\%  \\ \hline
LR & 54.83\% & 46.93\% \\ \hline
SVM & 51.52\% & 43.64\% \\ \hline
\end{tabular}
\end{table}

\vspace{0.1cm}
\noindent \textbf{Impact of attack dataset diversity.}
As shown in the previous experiment, the DNN model performs well when trained with all attack datasets.
Then a natural question is that if the DNN model is trained on a subset of attack datasets, will it achieve good performance when tested with unknown attacks?
In this experiment, we study the impact of attack dataset diversity on D3, and the results are provided in Table~\ref{table:attack_dataset_diversity}. 



As we can see, the performance of DNN depends on the attack datasets used for training. 
An interesting observation that when trained with A1, the classifier detects A3 with $P_d = 96.31\%$.
In contrast, when trained with A3, the classifier only detects A1 with $P_d = 21.80\%$.
In general, the model does not generalize to unknown attacks, suggesting the necessity of creating a diverse attack dataset for message classification.

\begin{table}[t]
\centering
\caption{Detection performance of D3 \rev{message classifier} when trained with different combinations of attack datasets}\label{table:attack_dataset_diversity}
\begin{tabular}{|c|c|c|c|c|}
\hline
\begin{tabular}[c]{@{}c@{}}Attack datasets\\ used for training\end{tabular} & $P_d$ of A1 & $P_d$ of A2 & $P_d$ of A3 & $P_{fa}$ \\ \hline
\{A1\} & \textbf{99.67}\% & 11.68\% & 96.31\% & 0.35\% \\ \hline
\{A2\} & 38.57\% & \textbf{99.07}\% &55.21\% & 0.40\% \\ \hline
\{A3\} & 21.80\% & 0.30\% & \textbf{100.00}\% & 0.03\% \\ \hline
\{A1, A2\} & \textbf{99.03}\% & \textbf{98.85}\% & 96.28\% & 0.59\% \\ \hline
\{A1, A3\} & \textbf{99.88}\% & 13.59\% & \textbf{100.0}\% & 0.70\% \\ \hline
\{A2, A3\} & 72.91\% & \textbf{98.94}\% & \textbf{99.97}\% & 0.27\% \\ \hline
\{A1, A2, A3\} & \textbf{99.36}\% & \textbf{98.34}\% & \textbf{100.00}\% & 0.43\% \\ \hline
\end{tabular}
\end{table}

\subsection{Performance of Aircraft Classification}
In this section, we evaluate the DNN-based aircraft classifier. 
We then study the impact of the size of the training set and the number of aircraft on the classification performance. 

\vspace{0.1cm}
\noindent \textbf{Setup.} 
We split the data into three portions: 60\% for training, 20\% for validation, and 20\% for testing. 
We post-processed the collected data and obtained a dataset that consists of 238 aircraft with 1000 messages for each aircraft, totalling 238,000 messages.
Each message has a total of 240 phase values (features) computed from the IQ samples.
Since DNN dominates other machine learning techniques (Section~\ref{sec:message_classification_performance}), we focus on DNN in this experiment.
Specifically, we consider five DNN models with various complexity: 
\begin{enumerate}
    \item M1: one hidden layer with 512 nodes,
    \item M2: one advanced hidden layer with 1024 nodes,
    \item M3: two hidden layers with 512 nodes per layer, and 
    \item M4: three hidden layers with 512 nodes per layer.
    \item \shaun{M5: a fine-tuned model with two hidden layers (512 and 256 nodes), batch normalization \cite{ioffe2015batch} and 200 epochs.}
\end{enumerate}
The parameters in Table~\ref{table:DNN_parameters} are used for constructing DNNs.

\vspace{0.1cm}
\noindent \textbf{Metrics.}
We adopt the following standard metrics for multi-class classification: \textit{precision}, \textit{recall}, and \textit{F-score} \cite{sokolova2009systematic}. 
Let $\text{TP}_i$ be the \textit{true positive rate} at which Aircraft $i$'s messages are correctly identified from Aircraft $i$.
Similarly, the \textit{false positive rate} $\text{FPR}_i$ and the \textit{false negative rate} $\text{FNR}_i$ are the rates of incorrectly accepted and incorrectly rejected identifications for Aircraft $i$, respectively.
Hence, we have
\begin{align*}
    \text{Precision: } \text{Pr}_i &= \frac{\text{TPR}_i}{\text{TPR}_i + \text{FPR}_i},\\
    \text{Recall: } \text{Re}_i &= \frac{\text{TPR}_i}{\text{TPR}_i + \text{FNR}_i},\\
    \text{F-Score: } \text{F-Score}_i &= \frac{2 \times \text{Pr}_i \times \text{Re}_i}{\text{Pr}_i + \text{Re}_i}.
\end{align*}
Note that high precision and high recall imply low FPR and low FNR, respectively. 

To further quantify the overall performance of the classifier, we compute the macro-averaging precision (AvgPr) and recall (AvgRe) by averaging the $\text{Pr}_i$'s and $\text{Re}_i$'s over all aircraft, and the average F-score as 
\begin{equation*}
    \text{AvgF-Score} = \frac{2 \times \text{AvgPr} \times \text{AvgRe}}{\text{AvgPr} + \text{AvgRe}}.
\end{equation*}
In addition, the accuracy metric is often used to measure the overall effectiveness of a classifier, defined as the percentage of messages that are correctly classified.

\vspace{0.1cm}
\noindent \textbf{Results.}
The results are provided in Table~\ref{table:DNN_performance}. 
Compared to M1 that achieves an average F-score of $93.64\%$ and an accuracy of $93.60\%$, increasing the model complexity by adding more nodes or one more layer (e.g., M2 and M3) improves the overall performance. 
\shaun{Keeping adding more layers (e.g., M4) does not necessarily result in  performance gain.
But with careful tuning of hyper-parameters and longer training time, a fine-tuned model like M5 can achieve an average F-score of $96.68\%$ and an accuracy of $96.66\%$.}


\begin{table}[t!]
\centering
\caption{Performance of DNN-based aircraft classifiers}
\label{table:DNN_performance}
\begin{tabular}{|c|c|c|c|c|}
\hline
Model & AvgPr & AvgRe & AvgF-Score & Accuracy \\ \hline
M1 & 93.69\% & 93.60\% & 93.64\% & 93.60\% \\ \hline
M2 & 94.62\% & 94.51\% & 94.57\% & 94.51\% \\ \hline
M3 & 94.88\% & 94.68\% & 94.78\% & 94.68\% \\ \hline
M4 & 94.52\% & 94.28\% & 94.40\% & 94.28\% \\ \hline
M5 & \textbf{96.69}\% & \textbf{96.66}\% & \textbf{96.68}\% & \textbf{96.66}\% \\ \hline
\end{tabular}
\end{table}

\begin{figure}[t!]
\centering
\includegraphics[width=1.0\columnwidth]{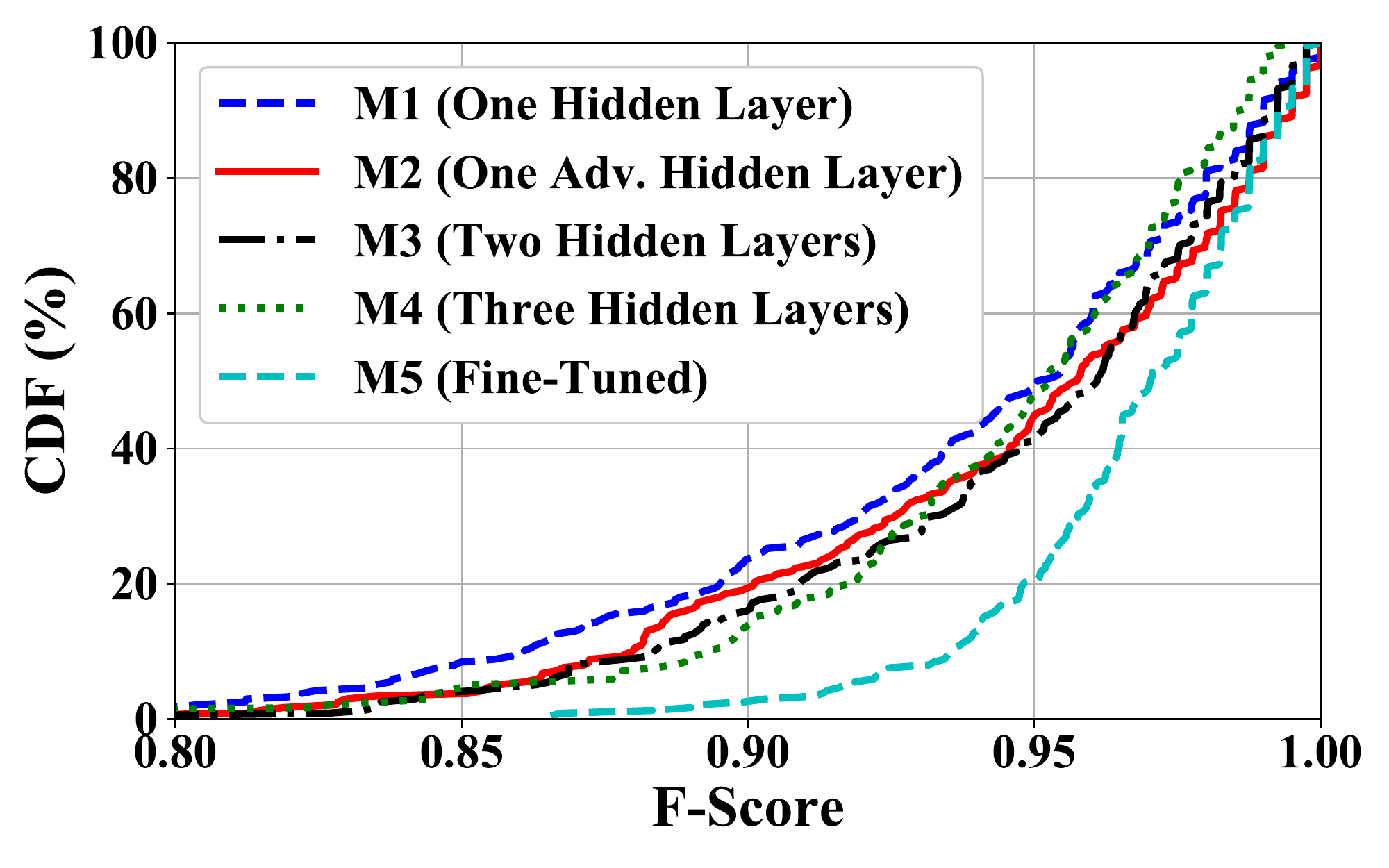}
\caption{Empirical CDF of F-scores of DNN-based \rev{aircraft} classifiers. }
\label{fig:F_score_CDF}
\end{figure}

Fig.~\ref{fig:F_score_CDF} plots the empirical cumulative distribution function (CDF) of the F-scores.
We can see that over $50\%$ of aircraft have a F-score of $95\%+$ for all models.
In addition, increasing model complexity tends to improve the F-score of individual aircraft.
For example, there are $23.53\%$ of aircraft with a F-score lower than $90\%$ with M1, and this ratio drops to $15.97\%$ and $13.87\%$ with M3 and M4, respectively. 
\shaun{For the fine-tuned model, this ratio is as low as $2.5\%$.}

\begin{figure}[t]
\centering
\includegraphics[trim={.5cm 0cm 4cm 2.5cm},clip,width=1\columnwidth]{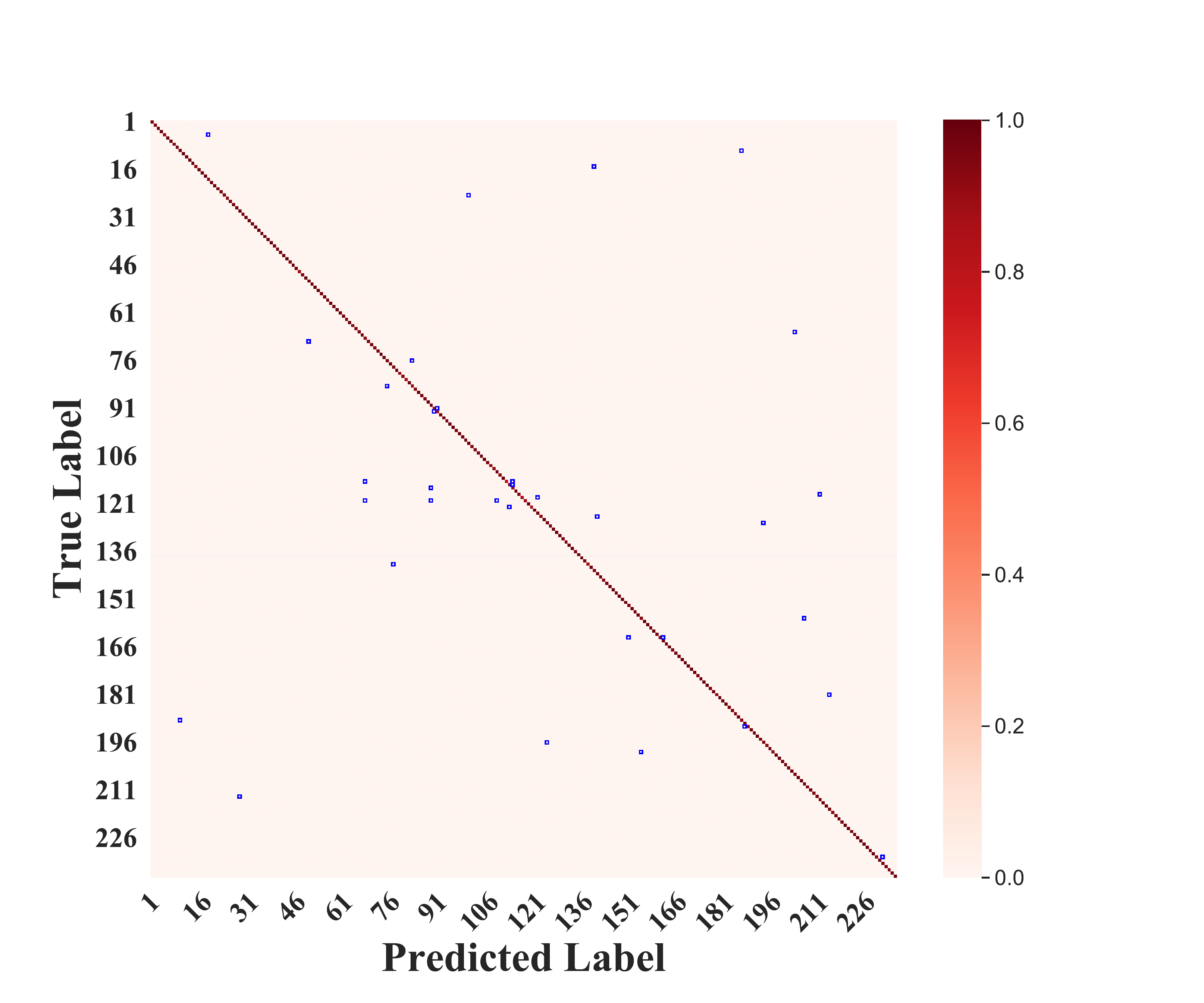}
\caption{Normalized confusion matrix of \shaun{the fine-tuned \rev{aircraft classifier} (M5)}.
There are $33$ non-diagonal entries larger than or equal to $0.02$ (blue squares).}
\label{fig:confusion_matrix}
\end{figure}

Fig.~\ref{fig:confusion_matrix} plots the confusion matrix of M5, in which the $(i,j)$-th entry represents the probability of Aircraft $i$'s messages incorrectly classified as Aircraft $j$'s. 
With \shaun{$96.66\%$} messages correctly identified, there exist only \shaun{$33$} non-diagonal entries (out of $56644$) with a rate larger than or equal to $2\%$ ($4$ misclassified messages). 
The largest misclassification rate is \shaun{$8.0\%$} between Aircraft 91 and 92, suggesting similarity in their phase patterns. 

\begin{figure}[t!]
\centering
\includegraphics[width=1.0\columnwidth]{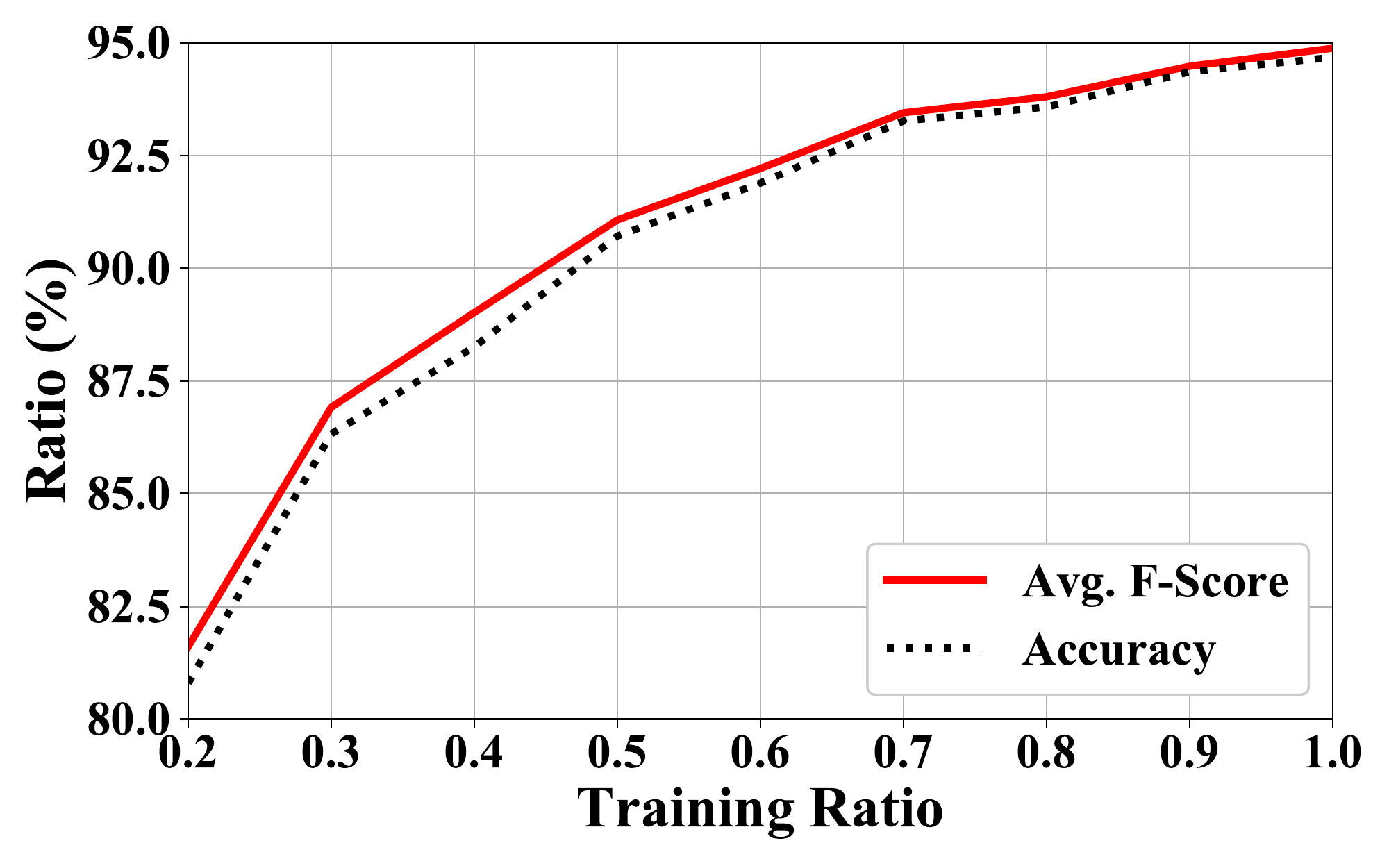}
\caption{Performance \rev{of the aircraft classifier (M3)} in terms of average F-score and accuracy as a function of the training ratio. The performance improves with more training samples and tends to saturate after a certain point.}
\label{fig:impact_of_training_size}
\end{figure}

\vspace{0.1cm}
\noindent \textbf{Impact of the size of the training set.}
In order to understand how the size of the training set affects the DNN performance, we vary the training ratio from $0.2$ to $1.0$ with a step of $0.1$ and plot the average F-score and accuracy of M3.
As shown in Fig.~\ref{fig:impact_of_training_size}, with only $20\%$ of the original training set (i.e., $120$ messages per aircraft), the classifier achieves an accuracy of over $80\%$. 
While the overall performance increases as the training ratio increases, the performance gain of more training samples tends to gradually decrease. 
\shaun{Similar trends are observed with other models and thus not reported due to the space limit.}

\begin{figure}[t!]
\centering
\includegraphics[width=1.0\columnwidth]{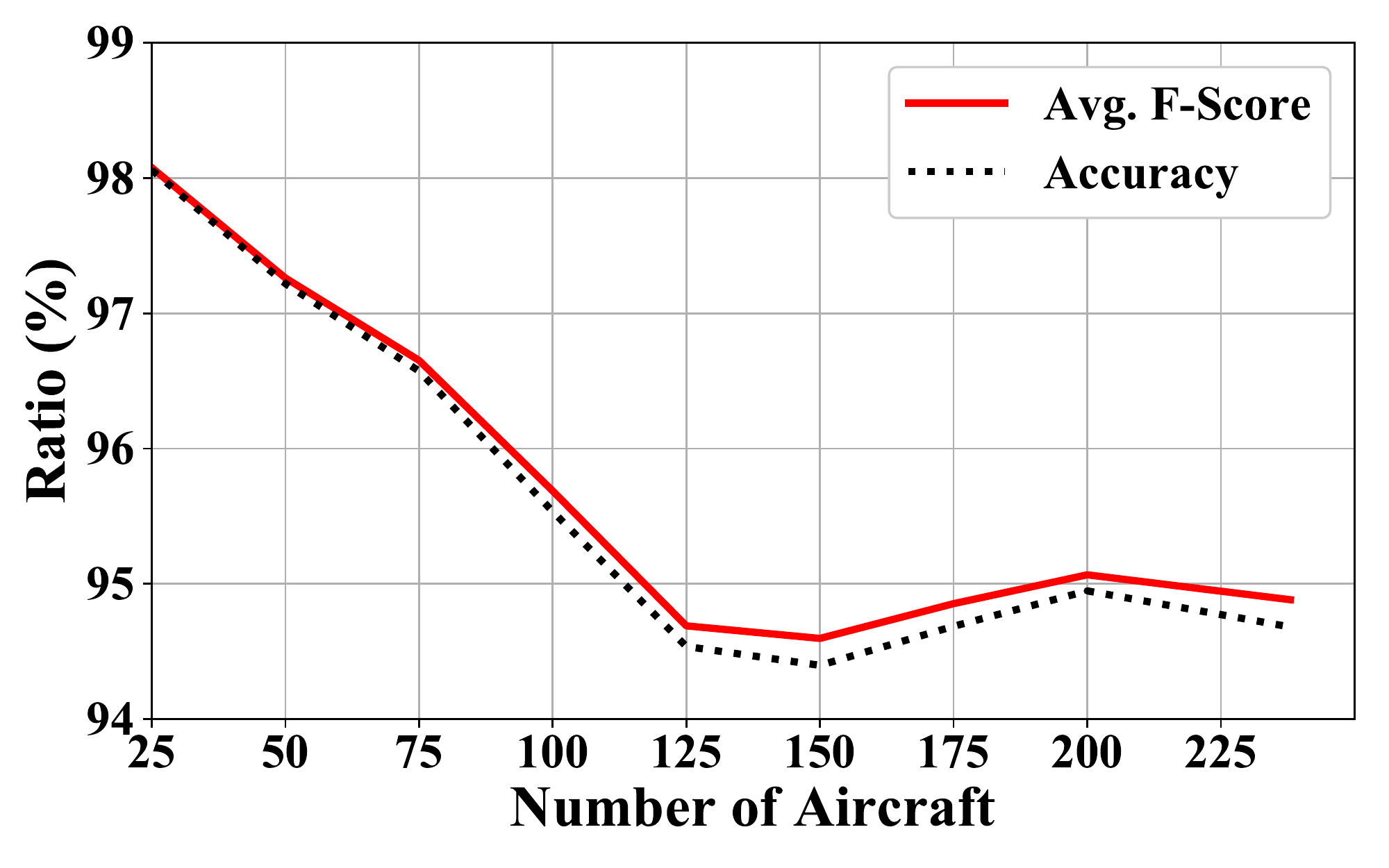}
\caption{Performance \rev{of the aircraft classifier (M3)} in terms of average F-score and accuracy as a function of the number of aircraft (classes).
When the number of aircraft increases, the performance slightly drops to around $95\%$. 
}
\label{fig:impact_of_num_of_aircraft}
\end{figure}

\vspace{0.1cm}
\noindent \textbf{Impact of the number of aircraft.}
In this experiment, we vary the number of aircraft from $25$ to $238$ with a step of $25$ and plot the performance of M3, as shown in Fig.~\ref{fig:impact_of_num_of_aircraft}. 
With only $25$ aircraft, the classification accuracy is as high as $98.06\%$. 
With more aircraft, the performance drops and comes to a level of around $95\%$. 


\balance
\section{Conclusion and Future Work}
\label{sec:conclusion}
In this paper, we proposed SODA to detect ADS-B spoofing attacks using DNN, which consists of a message classifier and an aircraft classifier. 
The message classifier takes IQ samples as input and detects malicious messages from the SDR-based ground attacker, including the message/IQ data replay attack and the ghost aircraft injection attack. 
The aircraft classifier takes the phases of received messages as input and detects spoofed messages by comparing the claimed ICAO address against the predicted ICAO address. 
We collected real-world ADS-B messages and generated realistic attack data using our ADS-B testbed through OTA experiments in an anechoic chamber. 
Our experimental results show that the SODA detects ground-based spoofing attacks with a detection probability of $99.34\%$ and a false alarm probability of $0.43\%$.
It detects aircraft spoofing attacks with an average F-score of $96.68\%$ and an accuracy of $96.66\%$. 

\rev{As future work, we will study the scalability of SODA and explore the use of clustering to identify aircraft with similar features and detect ADS-B spoofing attacks launched by aircraft in different clusters.
We will also perform post-processing of the IQ samples to extract more stable features (e.g., frequency offsets) to improve the robustness of SODA against various factors, including radio propagation, receiver characteristics, and measurement noise.}



\section*{Acknowledgments}
This work was supported in part by NSF grant CNS-1446866, ONR grants N00014-16-1-2710 and N00014-17-1-2946, and ARO grant W911NF-16-1-0485. 
Views and conclusions expressed are that of the authors and not be interpreted as that of the NSF, ONR or ARO.
\rev{This work was also supported by a grant of the Italian Presidency of the Council of Ministers.}

\bibliographystyle{IEEEtran}
\bibliography{./references}

\end{document}